\documentstyle[prl,aps,epsfig,multicol,amssymb]{revtex}
\begin{document}
\draft

\title{Can  Polymer Coils be modeled  as ``Soft Colloids''?}
\author{A.A. Louis$^1$, P.G. Bolhuis$^1$, J.P. Hansen$^1$, and
E.J. Meijer$^2$} \address{$^1$Department of Chemistry, Lensfield Rd,
Cambridge CB2 1EW, UK} \address{$^2$Department of Chemical
Engineering, University of Amsterdam, Nieuwe Achtergracht 166, NL-1018
WV Amsterdam, Netherlands.}
  \date{\today} \maketitle
\begin{abstract}
\noindent 
We map dilute or semi-dilute solutions of non-intersecting polymer
chains onto a fluid of ``soft'' particles interacting via a
concentration dependent effective pair potential, by inverting the
pair distribution function of the centers of mass of the initial
polymer chains.  A similar inversion is used to derive an effective
wall-polymer potential; these potentials are combined to successfully
reproduce the calculated exact depletion interaction induced by
non-intersecting polymers between two walls.  The mapping opens up the
possibility of large-scale simulations of polymer solutions in complex
geometries.
\end{abstract}
\vspace*{-0.2cm}
\pacs{61.25.H,61.20.Gy,82.70Dd}
\vspace*{-0.5cm}

\begin{multicols}{2}
A statistical description of polymer solutions in complex geometries,
such as the colloid-polymer mixtures which have recently received much
experimental attention\cite{Ilet95,Verm98,Poon99}, generally relies on
a nanometer scale segment representation of the polymer coils, a
computationally very demanding task except in the special case of
ideal (non-intersecting) polymers obeying Gaussian
statistics\cite{Meij94}.  This obviously follows from the fact that,
although the colloidal particles may reasonably be modeled by hard
impenetrable spheres or other complex shapes lacking internal
structure, each polymer coil involves $L$ segments which must satisfy
a non-intersection constraint. It thus appears  natural to attempt a
mesoscale coarse-graining, whereby polymer coils interact via
effective pair potentials acting between their centers of mass (CM).
Since polymers can interpenetrate, the effective potential $\beta v(r)$ is
expected to be soft, with a range of the order of the radius of
gyration $R_g$ of individual coils.  Such a coarse-grained description
has been a long-time goal in the statistical mechanics of polymer
solutions, dating back to the first attempts by Flory and
Krigbaum\cite{Flor50} who employed mean-field theory to find an
interaction for which the strength at overlap scales as: $\beta v(r$$=$$0) \sim
L^{0.2}$.  Later, scaling arguments\cite{Gros81}, field-theoretical
renormalization group calculations\cite{Krug89}, and
simulations\cite{Daut94} confirmed that the range of the interaction
between two isolated polymer coils is of order $R_g$, but found that
in the scaling limit the strength $\beta v(r$$=$$0)$ is independent of $L$ and
of order $k_BT$.

In this letter, we show that a meaningful ``soft colloid'' picture of
polymer coils may be built on a coherent ``first principles''
statistical mechanical foundation.  We derive both the effective
wall-polymer CM interaction $\beta \phi(z)$, and the ``best'' local
effective pair-potential $\beta v(r)$ between polymer CM's for {\em finite}
polymer concentrations. These potentials are then applied to simulate
bulk polymer solutions, as well inhomogeneous polymers near a hard
wall and polymers confined between two parallel walls to extract the
effective depletion potential between plates.  The ``soft colloid''
approach turns out to be successful not only in the dilute regime but
also, perhaps more surprisingly, well into the semi-dilute regime.  A
related ``soft particle'' picture has been applied to polymer melts
and blends\cite{Mura98}, but the corresponding phenomenological
implementation differs substantially from the present ``first
principles'' approach.

\begin{figure}
\begin{center}
\epsfig{figure=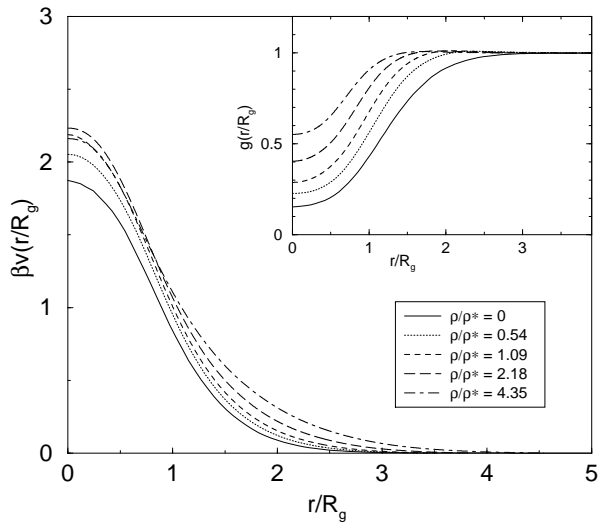,width=7cm,angle=270} 
\begin{minipage}{8cm}
\caption{\label{Fig1} The effective polymer CM pair potential
$\beta v(r/R_g)$ derived from an HNC inversion of $g(r/R_g)$ for different
densities.  The x-axis denotes $r/R_g$, where $R_g$ is the radius of
gyration of an isolated SAW polymer.  {\bf Inset:} The polymer CM pair
distribution function $g(r)$ calculated for $L$$=$$500$ SAW polymers and
used to generate $\beta v(r)$.  }
\end{minipage}
\end{center}
\end{figure}

We consider a popular model for polymers in a good
solvent\cite{Doi95}, namely $N$ excluded volume polymer chains of $L$
segments undergoing non-intersecting self avoiding walks (SAW) on a
simple cubic lattice of $M$ sites, with periodic boundary conditions.
The packing fraction is equal to the fraction of lattice sites
occupied by polymer segments, $c $$=$$ N\times L/M$, while the
concentration of polymer chains is $\rho $$=$$ c/L $$=$$ N/M$.  For a single
SAW chain, the radius of gyration $R_g \sim L^{\nu}$, where $\nu
\simeq 0.6$ is the Flory exponent\cite{Doi95}.  The overlap
concentration $\rho^*$, signaling the onset of the semi-dilute regime,
is such that $4 \pi \rho^*R_g^3/3 \simeq 1$, and hence $\rho^* \sim
L^{-3 \nu}$.  We have carried out MC simulations for chains of length
$L$$=$$100$ and $L$$=$$500$, and covered a range of concentrations up to
$\rho/\rho^* \sim 5$.  The pair distribution function $g(r)$ of the
centers of mass was computed for several concentrations; $g(r$$=$$0)$ is
always non-zero, thus confirming the ``softness'' of the effective
pair potential $\beta v(r)$.  The latter was then derived from $g(r)$ by an
inversion procedure based on the hyperneted-chain approximation (HNC) closure relation\cite{Hans86}:
\begin{equation}\label{eq1}
g(r) = \exp \{ - \beta v(r) + g(r) -c(r) -1 \},
\end{equation}
where $\beta$$=$$1/k_BT$, while $c(r)$ is the direct pair correlation
function, related to $g(r)$ by the Ornstein-Zernike (OZ)
relation\cite{Hans86}. To any given $g(r)$ and density there
corresponds a {\em unique} effective pair potential $\beta v(r)$,
capable of reproducing the input $g(r)$, {\em irrespective of the
underlying many-body interactions} in the system\cite{Hend74}; in a
variational sense this $\beta v(r)$ provides the ``best'' pair
representation of the true interactions\cite{Reat86}, and leads back
to the true thermodynamics via the compressibility
relation\cite{Hans86}.  While the simple HNC inversion procedure would
be inadequate for dense fluids of hard core particles, where more
sophisticated closures or iterative procedures are
required\cite{Reat86}, we are able to demonstrate the consistency of
the HNC inversion in the present case\cite{Bolh00}.  If the resulting
effective $\beta v(r)$, examples of which are shown in
Fig.~\ref{Fig1}, are used directly in MC simulations, the calculated
``exact'' $g(r)$ for this effective representation coincides within
statistical errors with the $g(r)$ derived from the simulation of the
full initial polymer segment model.  In fact, the HNC closure turns
out to be quasi-exact when applied to the simple Gaussian
model\cite{Stil76} whereby particles interact via the potential $\beta
v(r)$$=$$\epsilon \exp [ - \alpha \left( r/R_g \right)^2]$, which
yields a reasonable fit to the effective pair potentials shown in
Fig.~\ref{Fig1}.  Even the much cruder random phase approximation
closure, $c(r)$$=$$ -\beta v(r)$, yields semi-quantitatively accurate
results in the regime of interest\cite{Loui00a,Lang00}.  Careful
inspection of Fig.~\ref{Fig1} reveals that the effective pair
potential is not very sensitive to the polymer concentration.  The
value at $r$$=$$0$ first increases slightly with $\rho$, before
decreasing again at the highest concentration.  More strikingly, and
perhaps not surprisingly, the range of $\beta v(r)$ increases with
$\rho$.  The effective potential becomes slightly negative
$({\cal{O}}(10^{-3} k_BT))$ for $r/R_g \gtrsim 3 $ at the higher
concentrations.  \vglue -0.6cm
\vglue -0.65cm
 The properties of soft-core fluids are significantly different from
their hard-core counterparts.  For example, for potentials of the type
shown in Fig.~(\ref{Fig1}), the pressure is very well described by
$\beta P$$=$$ \rho + 1/2\beta \hat{V}(0) \rho^2$ over the entire density
range\cite{Loui00a,Lang00}.  Here $\hat{V}(0)$ is the Fourier
transform of the potential, at $k$$=$$0$.  Our observation that potentials
become slightly longer ranged at higher densities implies that the
pressure scales with an exponent slightly higher than $2$, so that the
equation of state (e.o.s.) is at consistent with the well-known
$\rho^{2.25}$ law\cite{Doi95}.  At first sight it may seem surprising
that a two-body potential could reproduce the full e.o.s.\ without
explicit many body terms.  However, the effective potential we use is
{\em constructed} to reproduce the true thermodynamics through the
compressibility relation (ignoring small volume terms); the
relative insensitivity of $\beta v(r)$ to concentration implies that
many-body interactions are not very important\cite{Loui00a}.

This insensitivity to concentration makes it possible to apply the
effective potential appropriate for a given mean concentration to
inhomogeneous cases, where the local polymer concentration deviates
from the mean.  Such a situation occurs when a polymer solution is
confined by a hard wall.  Using the same explicit SAW polymer model in
MC simulations, we have computed the ``exact'' profiles $h(z)$$=$$
\rho(z)/\rho -1$, where $z$ denotes the perpendicular distance of the
polymer CM from the wall. \vglue - 0.6cm
\begin{figure}
\begin{center}
\epsfig{figure=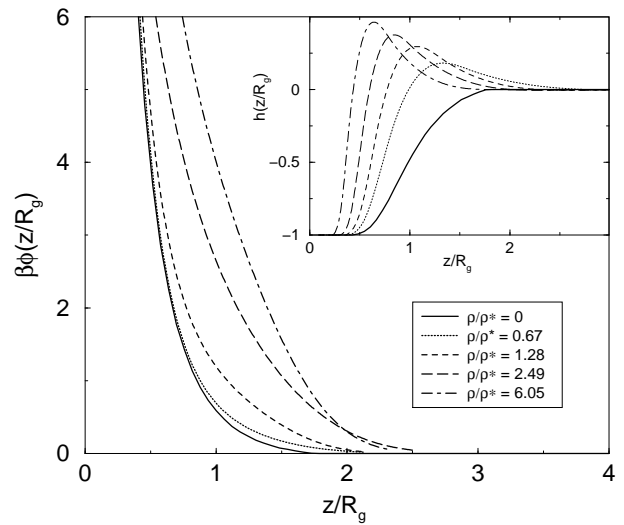,width=7cm,angle=270} 
\begin{minipage}{8cm}
\caption{\label{Fig2} The wall-polymer potential $\beta \phi(z/R_g)$
derived from an HNC inversion of $h(z)$.  {\bf Inset:} The
wall-polymer density profile $h(z) $$=$$ \rho(z)/\rho-1$ for different
densities.  The corresponding adsorptions $\Gamma$ are $0, 0.096,
0.132, 0.178$,and $0.248$ in units of $R_g^{-2}$ respectively.}
\end{minipage}
\end{center}
\end{figure} \vglue -0.6cm  Examples of $h(z)$ for several bulk
concentrations are shown in the inset of Fig.~\ref{Fig2}.  The
corresponding adsorptions $\Gamma$ are defined by:
\vspace*{-0.1cm}
\begin{equation}\label{eq2}
\Gamma = -\frac{\partial( \Omega^{ex}/A)}{\partial \mu} = \rho
\int_0^{\infty} h(z) dz,
\end{equation}
where $\Omega^{ex}/A$ is the excess grand potential per unit area,
$\rho$ the bulk concentration of the polymers and $\mu$ their chemical
potential.  From a knowledge of the concentration profile $\rho(z)$,
and  the bulk direct correlation function between polymers CM's
$c(r)$, one may extract an effective wall-polymer potential $\beta \phi(z)$
by combining the wall-polymer OZ relations\cite{Hans86} with the HNC
closure, resulting in:
 \vspace*{-0.3cm}
\begin{equation}\label{eq3}
\beta \phi (z) = \beta \phi^{MF}(z)
+ \rho \int d{\bf r'}
h(z') c(|{\bf r} - {\bf r'}|).
\end{equation}
The first term is the usual potential of mean force $\beta
\phi^{MF}(z) $$=$$ - \ln \left[\rho(z)/\rho\right]$, to which $\beta
\phi(z)$ would reduce in the $\rho \rightarrow 0$ limit, while the
second term arises from correlations between the polymer coils next to
the wall.  Using the $c(r)$ extracted from the earlier bulk
simulations of $g(r)$, together with Eq.~(\ref{eq3}), we are able to
extract $\beta \phi(z)$ from the density profiles.  Results for
various bulk concentrations are plotted in the Fig.~(\ref{Fig2}). 

The
range of the effective wall-polymer repulsion increases with
increasing concentration, while the density profiles actually move in
closer to the wall.  It is important to stress that the correlation
term considerably enhances the repulsion compared to the potential of
mean force.  We have tested the consistency of the inversion procedure
(which, to the best of our knowledge, has not been attempted before
for any wall/fluid interface) by using $\beta \phi(z)$, and the pair
potential $\beta v(r)$ for the appropriate bulk concentration, in MC
simulations based on these effective interactions (such simulations
are at least an order of magnitude faster than simulations of the
initial segment model).  The resulting concentration profile of the
effective ``soft colloids'' agrees to within statistical accuracy with
the initial $\rho(z)$ obtained from the detailed segment simulations,
and the corresponding adsorption $\Gamma$ differs by less than $1 \%$
from the exact value, thus demonstrating the adequacy of the ``soft
colloid'' representation of the interacting polymer coils.

\begin{figure}
\begin{center}
\epsfig{figure=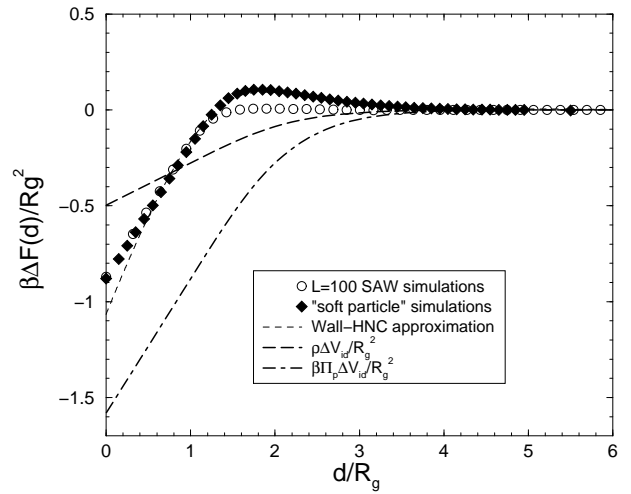,width=8cm,angle=0} 
\begin{minipage}{8cm}
\caption{\label{Fig3}
 Depletion free-energy $\beta \Delta F(d)/R_g^2$
for two plates separated by $d$. Circles are the ``exact'' MC
simulations of SAW polymers, the diamonds denote MC simulations of the
``soft particles'', the short dashed line denotes the wall-HNC
approximation of Eq.~(\protect\ref{eq4}).  The long-dashed and
dash-dotted lines denote the AO approximations mentioned in the text.}
\end{minipage}
\end{center}
\end{figure}

An even more severe test of this representation is provided by a
calculation of the depletion interaction between two hard walls
confining the polymers within a slit of width $d$.  Using direct
grand-canonical simulations of the full SAW polymer model, we computed
the osmotic pressure exerted by the polymer-coils on the walls; the
interaction free energy per unit area A, $\beta \Delta F/A$, is then
obtained by integrating the osmotic pressure calculated for different
values of the spacing $d$ between the walls.  These simulations are
extremely computer intensive, and were only be carried out for
$L$$=$$100$\cite{Meij00}.  In the ``soft colloid'' picture, the
interactions of the polymer CM's with each other, $\beta v(r)$, and
with a wall, $\beta \phi(z)$, are calculated once with the HNC
inversion procedures from the $g(r)$ and $\rho(z)$ of a full SAW
polymer simulation at the bulk density.  These are then used in
grand-canonical MC simulations of soft particles between two walls,
and in Fig.~\ref{Fig3} they are compared to the `exact''
grand-canonical MC simulations of $L$$=$$100$ SAW polymers (for
$\rho/\rho^*=0.95$) .  The results are in good agreement, but the
``soft colloid'' calculations are at least two orders of magnitude
faster.  Contrary to the more widely studied case of colloid-colloid
mixtures\cite{Gotz99}, the ``exact'' interaction exhibits no
significant repulsive barrier, whilst the ``soft colloid'' model leads
to a flat maximum; the corresponding barrier height is, however, very
small compared to the attractive minimum at contact, which agrees well
with the ``exact'' data, as does the slope of the attraction.  In
fact, the repulsive barrier does not increase significantly with
density\cite{Loui00a}, and its origin can be traced to our use of the
``potential overlap approximation'', namely that the interaction of
the soft particles with two parallel walls a distance $d$ apart can be
written as the sum of the two individual wall-particle
interactions. This is exact for simple liquids with true
intermolecular interactions, but not for polymers described by
effective potentials, even if the polymers are ideal\cite{Loui00a}.
For the sake of consistency, the MC simulations for the ``soft
colloid'' model were carried out with effective wall-polymer and
polymer-polymer potentials appropriate for $L$$=$$100$.  However, we
checked that the data obtained with effective interactions appropriate
for longer polymers ($L$$=$$500$), which cannot be easily handled
within the full segment model, are very close to the $L$$=$$100$
results, so we are confident that we are close to the scaling regime.

\vglue - 0.4cm
\vglue -0.5cm
In Fig.~\ref{Fig3} we also compare two results derived in the spirit
of the Asakura-Oosawa (AO) approximation\cite{Asak54}. The free-energy
difference $\beta \Delta F(z)$ is modeled by the density times the
exact depletion volume, $\Delta V_{id}(z)$, excluding one ideal
Gaussian polymer of size $R_g$, or by a popular phenomenological
improvement\cite{Ilet95}: $ \beta \Delta F(z) $$=$$ \beta \Pi_b \Delta
V_{id}(z)$, where $\Pi_b$ is the bulk osmotic pressure of the
interacting polymers.  Note that for the density under consideration,
these approximations are seen to be very poor, both as regards the
depth and the range of the depletion attraction. In fact, the range of
the depletion interaction for interacting polymer coils is
significantly reduced compared to the AO predictions, valid for ideal
polymers.  For low densities we find, as expected, that all the above
approaches converge\cite{Loui00a}.

These observations  can be understood within the ``soft colloid''
representation and the HNC approximation\cite{Atta91}, where the
interacting free energy per unit area   is given by: 
\vspace*{-0.2cm}
\begin{eqnarray}\label{eq4}
\frac{\beta \Delta F(z)}{A} & = & - \rho \int_{-\infty}^\infty h(s)
h(z -s) ds \nonumber \\ 
& +& \rho \int_{-\infty}^\infty h(z -s) \left
[ \beta \phi(s) - \beta \phi^{MF}(s) \right] ds.
\end{eqnarray}
Here $h(z) $$=$$ \rho(z)/\rho -1$ is the {\em single wall} density
profile, $\beta \phi(z)$ is the corresponding effective wall-polymer
potential, and $\beta \phi^{MF}(z)$ is the corresponding potential of mean
force.  The first term on the r.h.s.\ of Eq.~(\ref{eq4}) is the
density overlap approximation and would be the only-contribution in
the case of ideal (Gaussian) polymer coils.  The second term arises
from the correlations between coils; this 
dominates the first term in the semi-dilute regime $(\rho/\rho^* \geq
1)$. The standard AO approximation\cite{Asak54} may be derived from
Eq.~(\ref{eq4}) by replacing the density profile by a step-function of
width $R_g$ in the first term of Eq.~(\ref{eq4}) and neglecting the
correlation term.  In Fig.~\ref{Fig3} we compare the HNC approach of
Eq.~(\ref{eq4}) for the wall-wall interaction to the ``exact'' results
and the MC simulations of the ``soft colloids''.  As was found for the
homogeneous case and for the single wall, HNC works very well here,
demonstrating that knowledge of $\beta v(r)$ and $\beta \phi(z)$ quickly leads to
accurate predictions for the slit geometry, paving the way for the use
of integral equation techniques in other, more complex, geometries.

To summarize, the coarse-grained representation of polymer coils as
``soft'' colloids has been shown to be very reliable, yielding pair
distribution functions and concentration profiles which agree closely
with the results for the full SAW segment model, while being much more
efficient from a computational point of view.  Much of the success of
the coarse-graining lies in our finding that the ``best'' effective
pair potential between CM's of neighboring coils does not depend
strongly on polymer concentration, and is reasonably close to its
$\rho \rightarrow 0$ limit.  Similar conclusions were reached in
recent work on the phase-behavior of star-polymers, where the
$\rho\rightarrow 0$ limit of the pair potential was used to calculate
the phase-behavior at finite concentration\cite{Watz99}.  Our results
for the linear polymer case suggest that the full pair-potential for
star polymers may not be strongly concentration dependent, and that 
our approach could be used for star polymers in confined geometries.

Finally, we note that the ``soft colloid'' description is expected to
work best in complex geometries where the curvature is not too large
on the scale of $R_g$, such as colloid-polymer mixtures where the
colloid radius $R \leq R_g$. For such systems, the ``soft colloid''
model may now be used in large scale simulations or fluid integral
equations of polymers in complex geometries, such as the
structure\cite{Loui99a}, phase behavior\cite{Ilet95},
interactions\cite{Verm98}, and metastability\cite{Poon99} of colloid
polymer mixtures, which cannot be achieved with the detailed model of
non-intersecting polymer chains.

 AAL acknowledges support from the Isaac Newton Trust, Cambridge, PB
acknowledges support from the EPSRC under grant number GR$/$M88839,
EJM acknowledges support from the Royal Netherlands Academy of Arts
and Sciences.  We thank David Chandler, Daan Frenkel, Christos Likos,
Hartmut L\"{o}wen, and Patrick Warren for helpful discussions.

\end{multicols}

\vspace*{-.6cm}
\begin{thebibliography}{99}
\vspace*{-1.6cm}
\bibitem{Ilet95} S.M. Ilett, A. Orrock, W.C.K. Poon, and P.N. Pusey, 
Phys. Rev. E, {\bf 51}, 1344 (1995).
\bibitem{Verm98} R. Verma, J.C. Crocker, T.C. Lubensky, and A.G. Yodh,
Phys. Rev. Lett. {\bf 81}, 4004 (1998).
\bibitem{Poon99} W.C.K. Poon {\em et al.}, Phys. Rev. Lett. {\bf 83},
1239 (1999).
\bibitem{Meij94} E.J. Meijer and D. Frenkel, 
Phys. Rev. Lett. {\bf 67}, 1110 (1991); J. Chem. Phys.
{\bf 100}, 6873 (1994).
\bibitem{Flor50} P.J. Flory and W.R. Krigbaum, J. Chem. Phys. {\bf
18}, 1086 (1950).
\bibitem{Gros81} A.Y. Grosberg, P.G. Khalatur, and A.R. Khokhlov,
Makromol. Chem.,` Rapid Commun. {\bf 3}, 709 (1982).
\bibitem{Krug89} B.Kr\"{u}ger, L. Sch\"{a}fer, and A. Baumg\"{a}rtner,
J. Phys. France {\bf 50}, 319 (1989)
\bibitem{Daut94} J. Dautenhahn and Carol K. Hall, Macromolecules
 {\bf 27}, 5399 (1994) (and references therein).
\bibitem{Mura98} M. Murat and K. Kremer, J. Chem. Phys. {\bf 108},
4340 (1998).
\bibitem{Doi95} M. Doi,  {\em Introduction to Polymer Physics}
(Oxford University  Press, Oxford, 1995).
\bibitem{Hans86} J.P. Hansen and I.R. McDonald, {\em Theory of Simple
Liquids, 2nd Ed.} (Academic Press, London, 1986).
\bibitem{Hend74}R.L. Henderson, Phys. Lett. A. {\bf 49}, 197 (1974);
J.T. Chayes and L. Chayes, J. Stat. Phys., {\bf 36}, 471 (1984).
\bibitem{Reat86} L. Reatto, Phil. Mag. A {\bf 58}, 37 (1986);
L. Reatto, D. Levesque, and J.J. Weis, Phys. Rev. A. {\bf 33}, 3451 (1986).
\bibitem{Bolh00} To do the inversion we perform SAW simulations on a
($M$$=$$120\times120\times120$) cubic lattice with periodic
boundaries. The number $N$ of polymers ($L$$=$$500$ segments each) ranges
from $50$ to $500$. For the polymer-wall interaction we employed a
$M$$=$$160\times100\times100$ lattice with the wall perpendicular to the
long axis, P.G. Bolhuis, A.A Louis, and J.P.Hansen, {\em to be published}.
\bibitem{Stil76} F.H. Stillinger, J. Chem. Phys. {\bf 65}, 3968 (1976).
\bibitem{Loui00a} A.A. Louis, P. Bolhuis, and J.P. Hansen, 
preprint cond-mat/0007062; P.G. Bolhuis, A.A. Louis, J.P. Hansen and
 E.J.  Meijer, preprint. 
\bibitem{Lang00}A. Lang, C.N. Likos, M. Watzlawek, and H. L\"{o}wen,
 J. Phys.: Condens. Matter {\bf 12}, 5087 (2000)
\bibitem{Meij00}  For details of  the simulations:
 E.J. Meijer {\em et al.}, {\em to be published}.
\bibitem{Gotz99}Y. Mao, M.E. Cates, and H.N.W. Lekkerkerker, Physica A,
{\bf 222}, 10 (1995); B. G\"{o}tzelmann, R. Roth, S. Dietrich,
M. Dijkstra, and R. Evans, Europhys. Lett. {\bf 47}, 398 (1999);
S. Melchionna and J.P. Hansen, Phys. Chem. Chem. Phys. {\bf 2}, 3465
(2000). 
\bibitem{Asak54} S. Asakura and F. Oosawa, J. Chem. Phys. {\bf 22}, 1255
(1954), A. Vrij, Pure Appl. Chem. {\bf 48}, 471 (1976).
\bibitem{Atta91} P. Attard, D.R. B\'{e}rard, C.P. Ursenbach, and 
G.N. Patey, Phys. Rev. A {\bf 44}, 8224 (1991).
\bibitem{Watz99} M. Watzlawek, C.N. Likos, and H. L\"{o}wen,
Phys. Rev. Lett. {\bf 82}, 5289 (1999).
\bibitem{Loui99a} A. A. Louis, R. Finken, and J.P. Hansen,
Europhys. Lett. {\bf 46}, 741 (1999).
\end{thebibliography}
\end{document}